\documentclass{PoS}

\def\Black{} 

\title{{\vspace{-15mm} \normalsize\hfill{\small DESY 10-188}}\\[15mm]{\vspace{-18mm} \normalsize\hfill{\small SFB/CPP-10-104}}\\[15mm]{\vspace{-18mm} \normalsize\hfill{\small LPSC10165 }}\\[10mm]Low lying baryon spectrum with $N_f=2+1+1$ dynamical twisted quarks}
\ShortTitle{Low lying baryon spectrum with $N_f=2+1+1$ dynamical twisted quarks}

 \author{\speaker{Vincent Drach}\thanks{On behalf of the European
       Twisted Mass Collaboration}, Karl Jansen\\
NIC, DESY, Zeuthen, Platanenallee 6, D-15738 Zeuthen, Germany\\
Email \email{vincent.drach@desy.de,Karl.Jansen@desy.de}}
\author{Jaume Carbonell, Mauro Papinutto   \\  
     Laboratoire de Physique Subatomique et de Cosmologie,
       UJF, CNRS/IN2P3, INPG \\
        E-mail: \email{carbonel@lpsc.in2p3.fr,mauro.papinutto@lpsc.in2p3.fr}} 
\author{Constantia Alexandrou \\
  Department of Physics, University of Cyprus, P.O. Box 20537,
 1678 Nicosia, Cyprus\\
        E-mail: \email{alexand@ucy.ac.cy}}

\abstract{
\centerline{\includegraphics{Logo.pstex}}
We present first results on the octet and decuplet strange baryon spectrum with 
$N_f=2+1+1$ twisted mass quarks. 
We use an Osterwalder Seiler valence strange quark with a mass tuned to the kaon and compare 
the results with those obtained in the unitary setup.
This comparison allows to perform a first study of the lattice artefacts introduced by the mixed action approach. 
We investigate the effect of the strange and charm quarks in the sea by using two lattice 
spacings and comparing with preceding $N_f = 2$ twisted mass fermion calculations. 
}

\FullConference{The XXVIII International Symposium on Lattice Field Theory, Lattice2010\\
		June 14-19, 2010\\
		Villasimius, Italy}

\begin{document}

\section{Introduction}

The European Twisted Mass (ETM) collaboration  is pursuing its effort to generate $N_f=2+1+1$ configurations of  twisted mass fermions. 
This discretization of QCD has  the advantage of providing automatic $O(a)$ improvement  of all the observables once tuned at maximal twist~\cite{Chiarappa:2006ae,Baron:2010bv}. In this contribution we are interested in computing the masses of the baryon octet and decuplet.   
The inclusion of the strange and charm quarks in the sea  offers the possibility to investigate several effects in the baryon sector. 

First, given the large value of the charm quark masss, potentially large cutoff effects could appear in observables believed to be insensitive to the 
charm degrees of freedom.  We perform a preliminary analysis of the discretization effects using three lattice spacings. 
The non degenerate twisted mass fermion doublet used  in the heavy sector do not conserve the strange and charm quantum numbers at finite lattice spacing.  The effects of this symmetry breaking have already been investigated in~\cite{Baron:2010th} to determine the kaon and D meson masses. In order to study the spectrum of strange baryons, two approaches can nevertheless be followed: the mixed action setup using an action that conserves strangeness in the valence or the unitary setup using the sea action. Both methods have been followed to investigate unitarity violations. Finally, by comparing present results with our previous work using $N_f=2$ configurations in the partially quenched approximation~\cite{Alexandrou:2009qu}, we investigate the effect of the quenching of the strange quark.

\section{Lattice setup} \label{se:setup}

 In the gauge sector we use the Iwasaki gauge action~\cite{Iwasaki:1985we}. Using the same notations as in Ref.~\cite{Baron:2010bv}, the light doublet action reads :
\begin{equation}
  \label{eq:sl}
  S_l\ =\ a^4\sum_x\left\{ \bar\chi_l(x)\left[ D[U] + m_{0} +
    i\mu_l\gamma_5\tau_3\right]\chi_l(x)\right\}\, ,
\end{equation}
while the action in the heavy non degenerate doublet  is :
\begin{equation}
  \label{eq:sf}
  S_h\ =\ a^4\sum_x\left\{ \bar\chi_h(x)\left[ D[U] + m_{0} +
    i\mu_\sigma\gamma_5\tau_1 + \mu_\delta \tau_3 \right]\chi_h(x)\right\}\, ,
\end{equation}

\begin{table}[h!]
  \centering
  \begin{tabular*}{0.9\textwidth}{@{\extracolsep{\fill}}lcccccc}
    \hline\hline
    Ensemble & $\beta$ & $a\mu_l$ & $a\mu_\sigma$ & $a\mu_\delta$ & $(L/a)^3 \times T/a$\\
    \hline\hline
    A30.32&1.90&0.0030&0.150&0.190&$32^3 \times 64$\\
    A40.32&&0.0040&&&$32^3 \times 64$\\
    A40.24&&0.0040&&&$24^3 \times 48$\\
    A50.32&&0.0050&&&$32^3 \times 64$\\
    A60.24&&0.0060&&&$24^3 \times 48$\\
    A80.24&&0.0080&&&$24^3 \times 48$\\
    A100.24&&0.0100&&&$24^3 \times 48$\\
      \hline
    B25.32&1.95&0.0025&0.135&0.170&$32^3 \times 64$\\
    B35.32&&0.0035&&&$32^3 \times 64$\\
    B55.32&&0.0055&&&$32^3 \times 64$\\
    B75.32&&0.0075&&&$32^3 \times 64$\\
    B85.24&&0.0085&&&$24^3 \times 48$\\
    \hline
    D20.48&2.10&0.0020&0.120&0.1385&$48^3 \times 96$\\
    D30.48&&0.0030&&&$48^3 \times 96$\\
\hline
  \end{tabular*}
  \caption{Summary of the $N_{\rm f}=2+1+1$ ensembles used in this work.}
  \label{tab:ensembles_211}
\end{table}

The input parameters of the calculation  $(L/a,\beta,\mu_l,\mu_{\sigma},\mu_{\delta})$ are collected in Table \ref{tab:ensembles_211}. They span a pion mass
range from approximately 270 to 500 MeV and $m_{\rm{PS}} L \ge 3.4$ ( the lowest bound holds for ensembles B25.32 and A40.24).   Estimation of the lattice spacing from Ref.~\cite{Baron:2010bv} gives $a_{\beta=1.90} = 0.0861(4)$~fm  and  $a_{\beta=1.95} = 0.0778(4)$~fm. A preleminary estimate of the scale at $\beta=2.10$ gives $a_{\beta=2.10} \approx 0.060$~fm. \\

The masses are extracted from two-points correlators using the same interpolating fields as in our previous $N_f=2$ study~\cite{Alexandrou:2009qu}. Errors are estimated using the bootstrap method.

To improve the overlap of the interpolating fields with the ground states we apply Gaussian smearing to each quark fields defining $\tilde{q}({\bf x},t) = \sum_{\bf y} F({\bf x},{\bf y};U(t)) q({\bf y},t)$
using the gauge invariant smearing function  
\begin{equation} 
F({\bf x},{\bf y};U(t)) = (1+\alpha H)^ n({\bf x},{\bf y};U(t)),
\end{equation}
constructed from the hopping matrix,
$ 
H({\bf x},{\bf y};U(t))= \sum_{i=1}^3 \biggl( U_i({\bf x},t)\delta_{{\bf x,y}-i} +  U_i^\dagger({\bf x}-i,t)\delta_{{\bf x,y}+i}\biggr).
$
 Furthermore we apply APE smearing to the spatial links  that enter the hopping matrix.
The parameters for the Gaussian and APE smearing are given in Table~\ref{tab:smearing_parameters_nf2p1p1}.

\begin{table}[h!]
\begin{center}
\begin{tabular}{ccccc}
\hline
\hline
$\beta$ & $N_{\mathrm{Gaussian}}$ & $\alpha_{\mathrm{Gaussian}}$ & $N_{\mathrm{APE}}$ & $\alpha_{\mathrm{APE}}$\\
\hline
\hline
$1.90$  & $50$ & $4.0$ &  $20$ & $0.5$ \\
$1.95$  &  $50$ & $4.0$ &  $20$ & $0.5$ \\
$2.10$  &  $70$ & $4.0$ &  $20$ & $0.5$ \\
\hline
\end{tabular}
\end{center}
\caption{Smearing parameters used in this work.}
\label{tab:smearing_parameters_nf2p1p1}
\end{table}
Two approaches are followed to investigate strange baryons using our $N_f=2+1+1$ simulations.

 The first one consists in using a mixed action in order to avoid the mixing of the strange and charm quantum numbers. 
As in our $N_f=2$ study, we choose  Osterwalder-Seiler fermions in the valence to simulate the strange quark. They are defined by the following action:
\begin{equation}
S^{\mathrm{OS}} = a^4\sum_x  \bar{s}(x)\bigl(D[U] + m_{0} +i \mu_s \gamma_5  \bigr ) s(x)
\label{S_os}
\end{equation}
The parameter $a\mu_S$ has been tuned using the kaon mass calculated in the unitary theory on each gauge ensembles by imposing :
\begin{equation}
 m^{\mathrm{OS}}_{K}(a\mu_s,\beta,a\mu_l,a\mu_\sigma,a \mu_\delta) =   m_{K}^{\mathrm{unitary}}(\beta,a\mu_l,a\mu_\sigma,a \mu_\delta)
\end{equation}
In practice one performs the measurement of the baryon masses for several values of the bare strange quark mass $a\mu_s$ and interpolate linearly to the matched quark mass. All the results presented in the following are interpolated at the matched strange quark mass. \\

The second method is using the unitary theory as defined in Eq.~(\ref{eq:sf}). The measurements of the kaon and D meson masses in the unitary setup have been extensively discussed in Ref.~\cite{Baron:2010th}. In particular, one can show that performing a rotation to the physical basis assuming all the renormalization are equal to one (the so called pseudo physical basis) leads to a propagator that simulates a strange quark for large Euclidean time.  We use this strategy to extract the octet and decuplet of strange baryons.

Note that the two approaches differ by $\mathcal{O}(a^2)$ effects and that the difference in the results between the two approaches is an unitarity violation.

\section{Results in the light sector }

\begin{figure}[h]
\begin{minipage}[ht]{7.5cm}
\includegraphics[height=6cm,width=7.5cm]{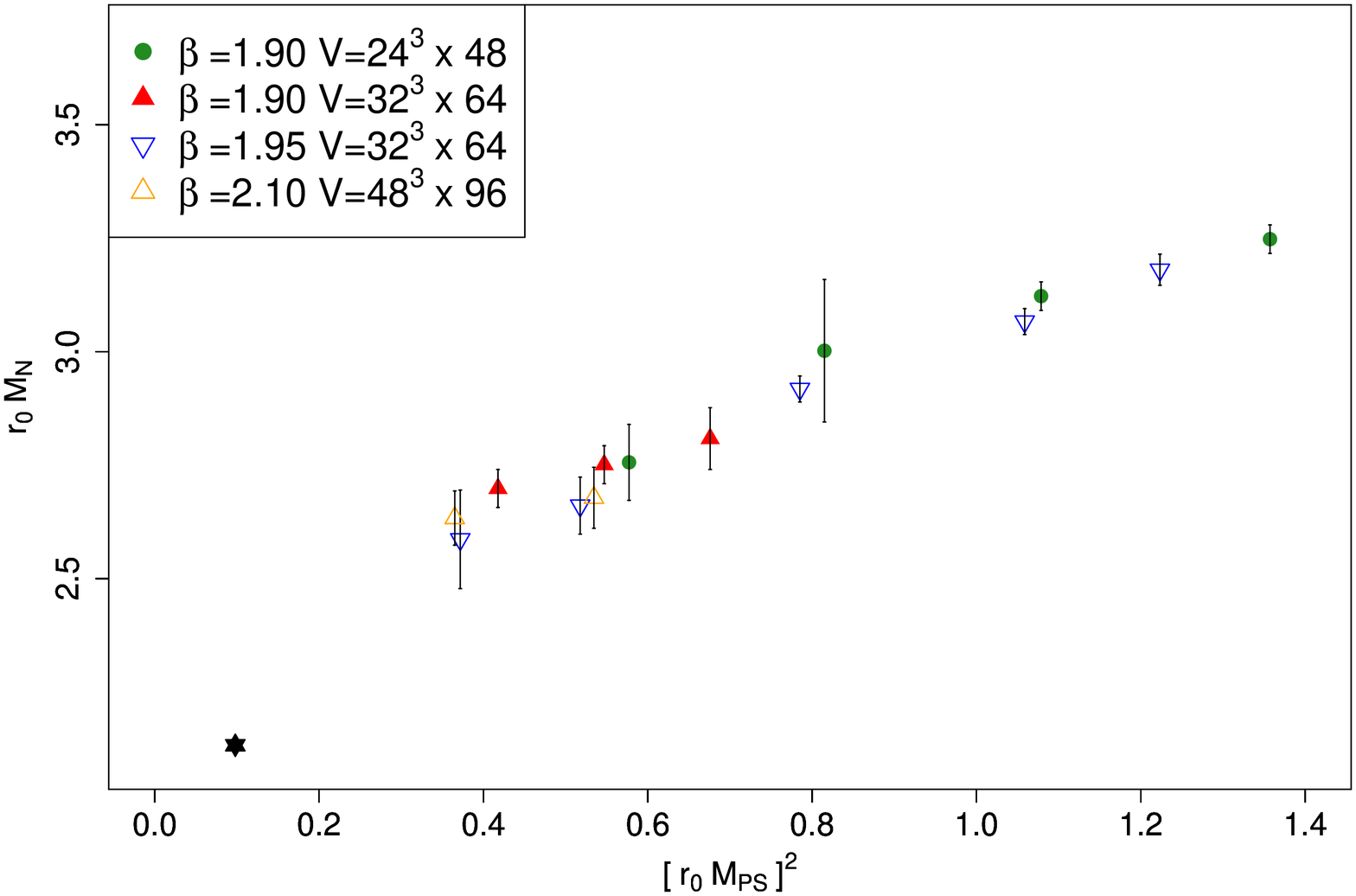}
\caption{Nucleon mass in units of $r_0$ as a function of $(r_0 m_{\rm{PS}})^2$. Physical point is indicated by a black star and
 the scale has been set using $r_0(\beta=1.95)=0.447(5)$~fm from \cite{Baron:2010bv} }
\label{fig:r0nuc_all}
\end{minipage}
\hspace{0.5cm}
\begin{minipage}[ht]{7.5cm}\vspace*{-0.9cm}
\includegraphics[height=6cm,width=7.5cm]{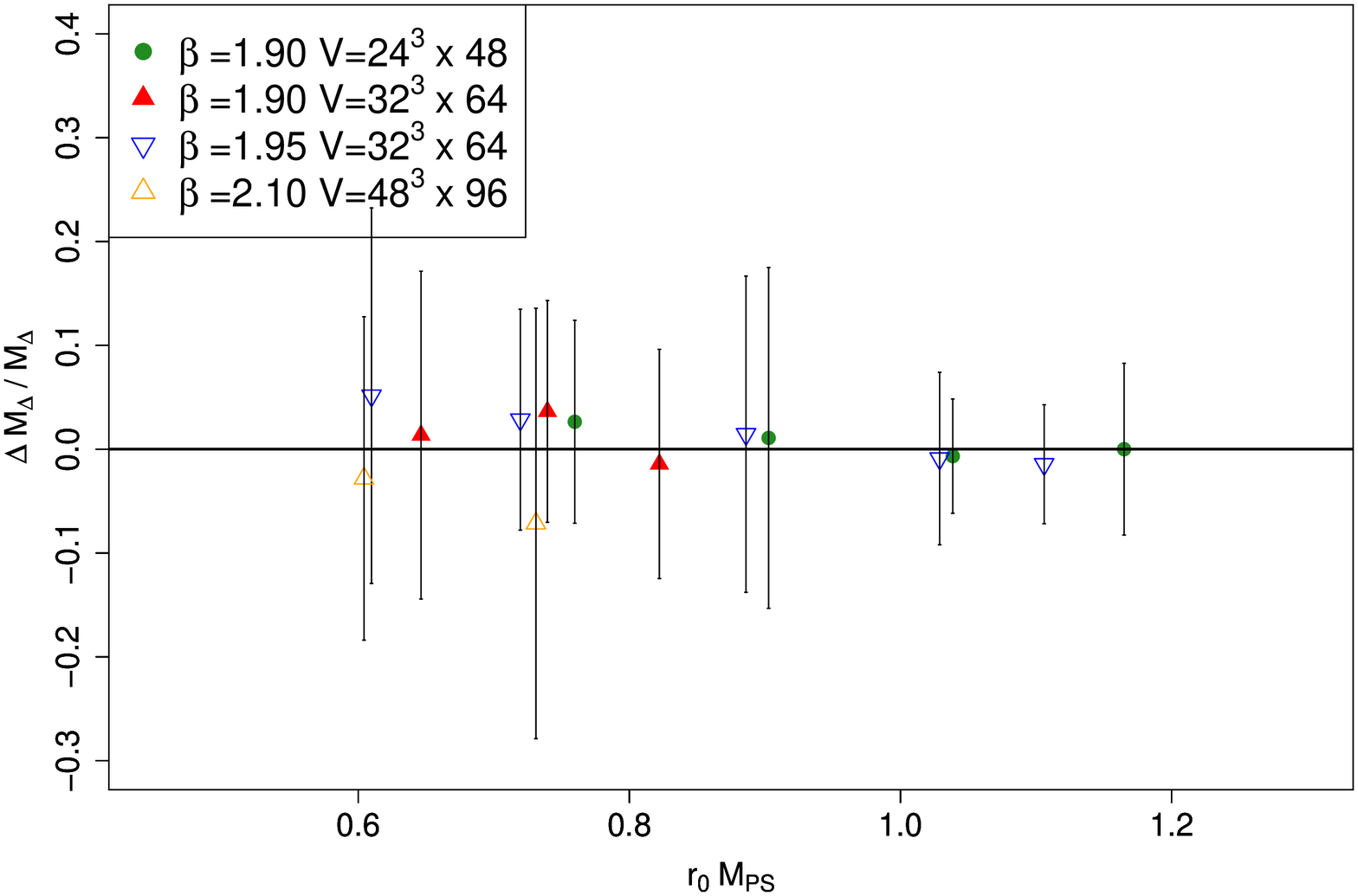}
\caption{Difference of the $\Delta^{++,-}$ and   a $\Delta^{+,0}$  masses divide by their average  as a function of $r_0m_{\rm{PS}} $}
\label{fig:isb_delta}
\end{minipage}
\end{figure}

We show in Fig.~\ref{fig:r0nuc_all}, our results for the nucleon mass in $r_0$ units  as a function of $(r_0 m_{\rm{PS}})^2$. The physical point is shown using  $r_0=0.447(5)$~fm from Ref.~\cite{Baron:2010bv}. In this plot as well as in all the following ones, a systematic error on the plateau determination has been added quadratically to the statistical error. The results obtained are consistent, indicating that finite volume and lattice spacing effects are  small.

Another important issue raised by twisted mass fermions, is the isospin symmetry breaking 
at finite lattice spacing. In the $N_f=2$ case we have shown in Ref.~\cite{Alexandrou:2008tn} that  in the $\Delta$ sector, isospin symmetry breaking manifests itself as a mass spliting between $\Delta^{++,-}$ and $\Delta^{+,0}$ states and  was found to be compatible with zero. 
The same conclusion holds for $N_f=2+1+1$ results as shown in Fig.~\ref{fig:isb_delta}, where we display 
the relative mass difference between the two states for three $\beta$ values as a function of $r_0 m_{\rm{PS}}$. 

Given that we have results for three lattice spacing, it is worthwhile to perform a combined fit of our results. To this aim we used the fitting function 
\begin{equation}
r_0 m_{N} = r_0 m_N^{(0)} +  \frac{4c^{1}_N}{r_0} (r_0  m_{\rm{ps}})^2  - \frac{3g_A^2}{16\pi f_{\pi}^2 r_0^2} (r_0 m_{\rm{PS}})^3  + C (\frac{a}{r_0})^2
\label{eq:xfit_Op3_nuc}
\end{equation}
where the parameter $m_N^{(0)}$ is the nucleon mass in the chiral limit, $c^{1}_N$ is related to the 
nucleon sigma term, $g_A$ is nucleon axial coupling, $f_{\pi}$ the pion decay constant and $C$ is a coefficient  that controls the discretization effects. 
In our fits, the values of $g_A$ and $f_{\pi}$ are fixed to their experimental values \textit{i.e} $g_A = 1.26$  and $f_{\pi} =0.1307$~GeV. 
This parametrization comes from an $\mathcal{O}(p^3)$ expansion in heavy baryon chiral perturbation theory (see Ref.~\cite{Alexandrou:2009qu} and reference therein for details). 
In order to take into account possible discretization effects a term proportional to $a^2$ has been added in Eq.~(\ref{eq:xfit_Op3_nuc}).

\begin{figure}[h!]
\begin{minipage}[ht]{7.5cm}
\includegraphics[height=6cm,width=7.5cm]{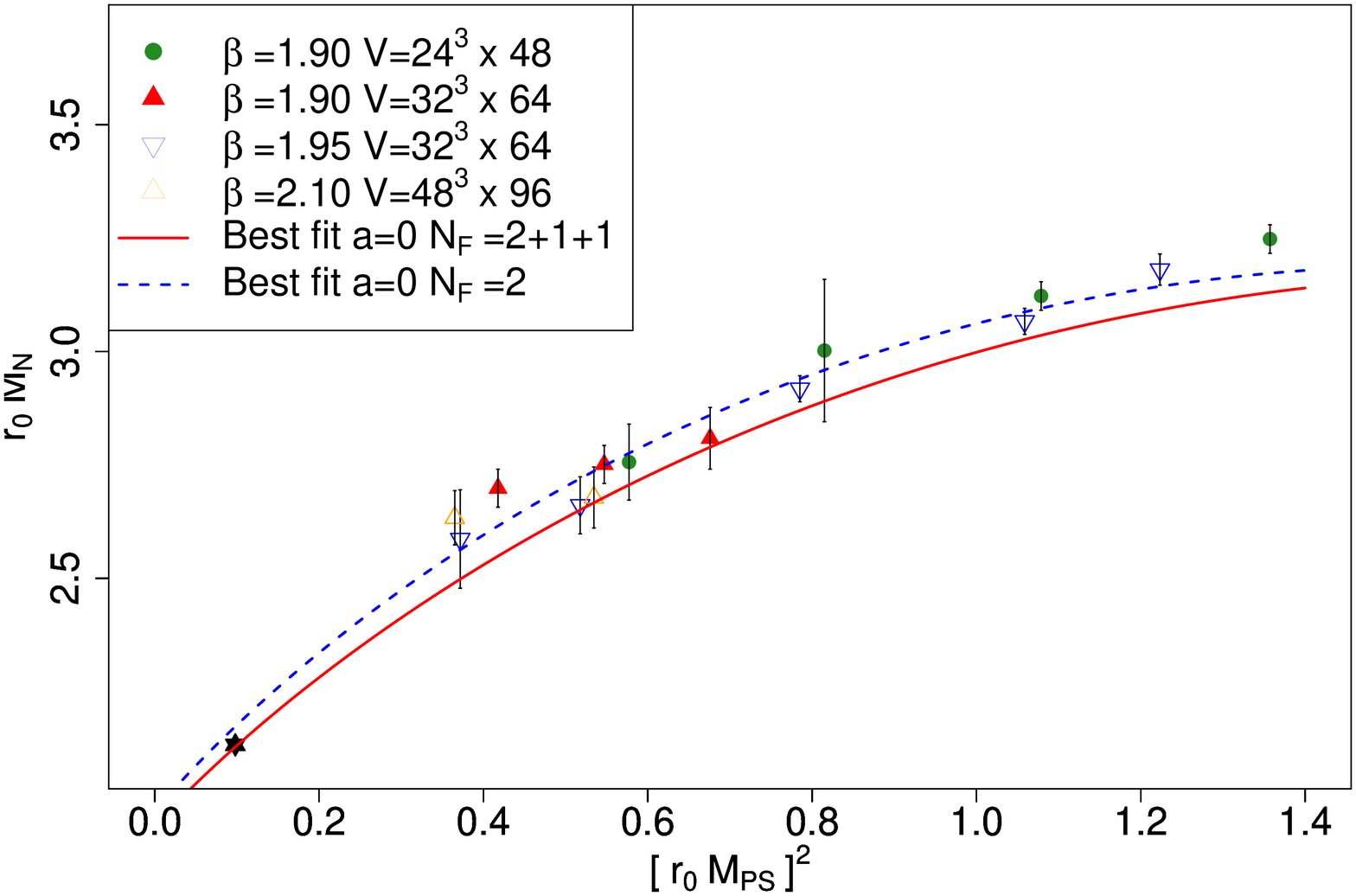}
\caption{Nucleon mass in $r_0$ units as a function of $(r_0 m_{\rm{PS}})^2$. The red curve comes from the combined fit using Eq.~(3.1) of all the $N_f=2+1+1$ data, setting $a=0$. The blue dotted lines correspond to the same results with $N_f=2$. The physical point is not included in the fit. }
\label{fig:xfit_global_r0nuc}
\end{minipage}
\hspace{0.5cm}
\begin{minipage}[ht]{7.5cm}\vspace*{-0.1cm}
\includegraphics[height=6cm,width=7.5cm]{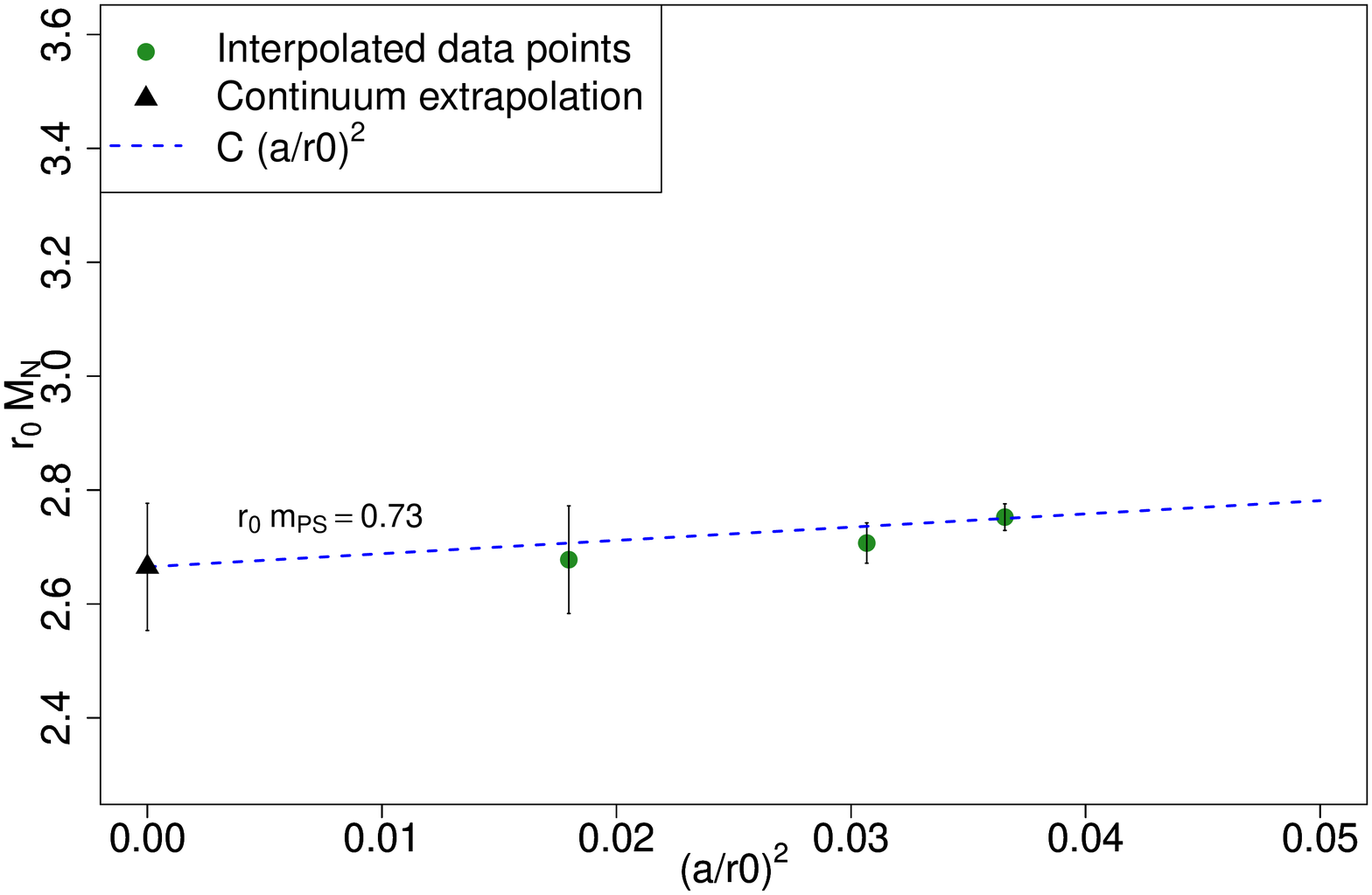} 
\caption{Nucleon mass as a function of $(a/r_0)^2$ for a fixed pion mass reference. The green dots are the nucleon masses interpolated at $r_0 m_{\rm{PS}} = 0.73$. The blue dotted line is obtained from our combined fit. The continuum extrapolation is shown by a black triangle.}
\label{fig:scaling_nuc}
\end{minipage}
\end{figure}

We show in Fig.~\ref{fig:xfit_global_r0nuc}, the nucleon mass as a function of $(r_0 m_{\rm{PS}})^2$.  The lattice results were fitted using Eq.~(\ref{eq:xfit_Op3_nuc}) and  the best parametrization extrapolated to the continuum (setting $a=0$ ) is 
denoted by a red curve. The fit has a $\chi^2_{\rm{d.o.f}} = 11.2/10$. The finite lattice spacing data are compatible with our estimation of the continuum behaviour, showing that the lattice spacing dependence is small.  
In order to assess for possible quenching effects of the strange quark we show with a blue dotted line the same fit performed using our $N_f=2$ data at three lattice spacings ($\chi^2_{\rm{d.o.f}} = 5.5/7$ ). The two  fits agree with each other,  indicating that, within errors, the nucleon mass is not sensitive to the presence of the strange quark in the sea. 

The scaling behaviour of the nucleon mass, for a fixed pion mass reference, is shown in Fig.~\ref{fig:scaling_nuc}. 
The value  $r_0 m_{\rm{PS}} = 0.73$  is in a region where we have data for the three lattice spacings. The finite lattice spacing points have been interpolated from the data to match the reference pion mass. The blue dotted line indicates the best fit parametrization of our results. Our estimation of the coefficient $C$  is compatible with zero, confirming that discretization effects are smaller than our statistical errors.

\section{Results in the strange sector }

\begin{figure}[h!]
\vspace{-.5cm}
\centering{\includegraphics[height=6cm,width=7.5cm]{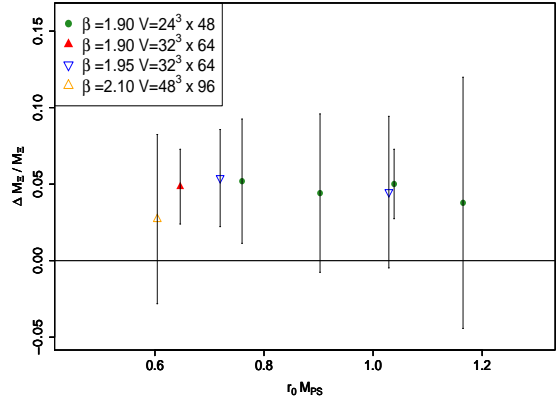} }
\vspace{-.2cm}
\caption{Mass splitting of the $\Xi$ as a function of $r_0 m_{\rm{PS}}$ for the three lattice spacings.}\label{fig:isb_xi}
\end{figure}

In our previous study on the strange baryons \cite{Alexandrou:2009qu} we have shown that the isospin breakingin was manifests itself in the $\Xi$
sector. We show in Fig.~\ref{fig:isb_xi} the mass splitting  for the $\Xi$ as a function of $r_0 m_{\rm{PS}}$ for the three lattice spacings. 
Most of our results are now  compatible with zero.

\begin{figure}[h!]
\vspace{-.4cm}
\begin{minipage}[ht]{7.5cm}
\includegraphics[height=6cm,width=7.5cm]{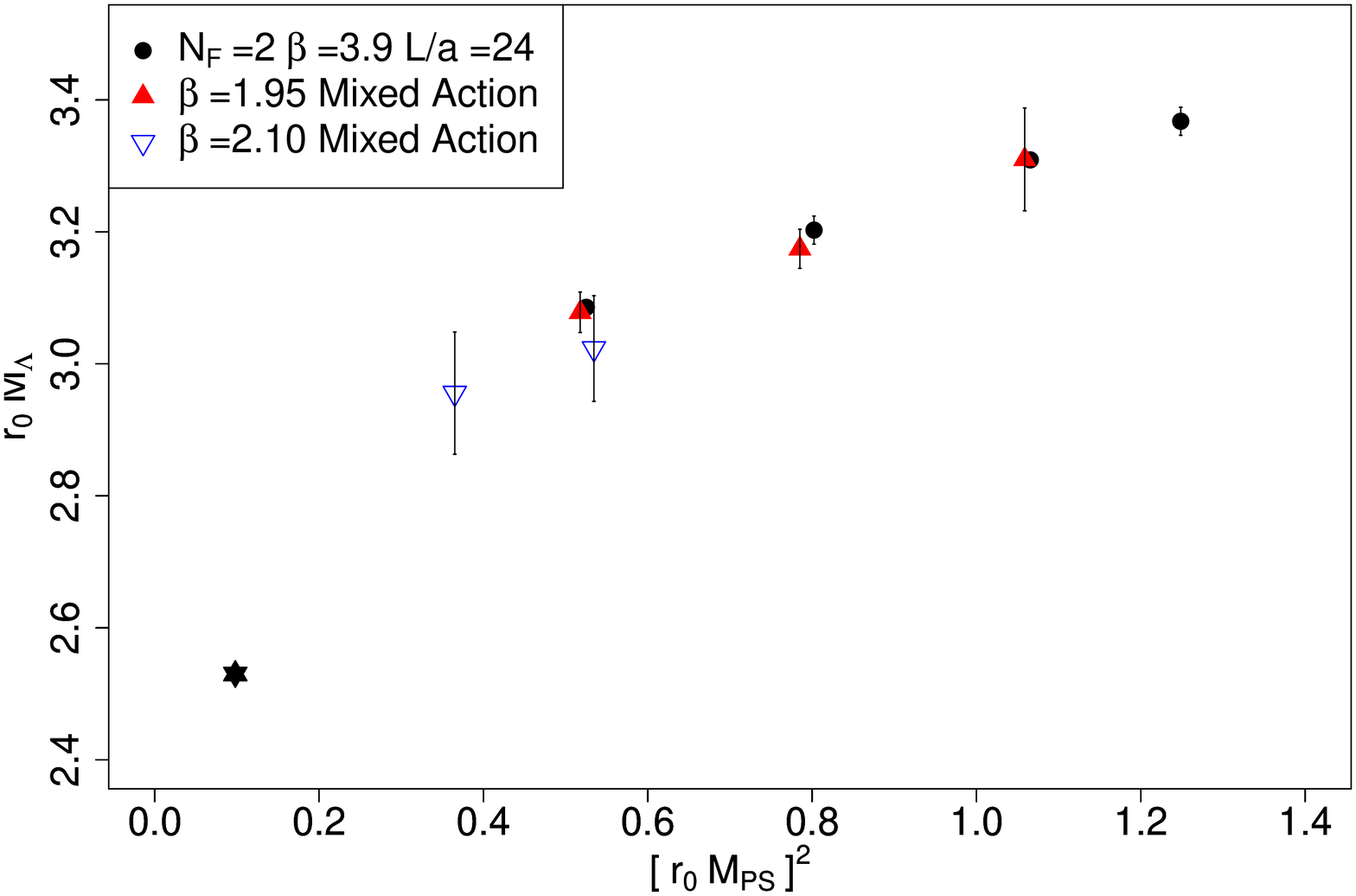}
\vspace{-.4cm}
\caption{$\Lambda$  mass as a function of $(r_0 m_{\rm{PS}})^2$ in the mixed action approach. For comparison $N_f=2$ data are shown with black dot. The physical point
 is indicated by  a black star. The scale was fixed using  $r_0$ from Ref.~\cite{Baron:2010bv}}
\label{fig:r0lambda}
\end{minipage}
\hspace{0.5cm}
\begin{minipage}[ht]{7.5cm}\vspace*{-0.1cm}
\includegraphics[height=6cm,width=7.5cm]{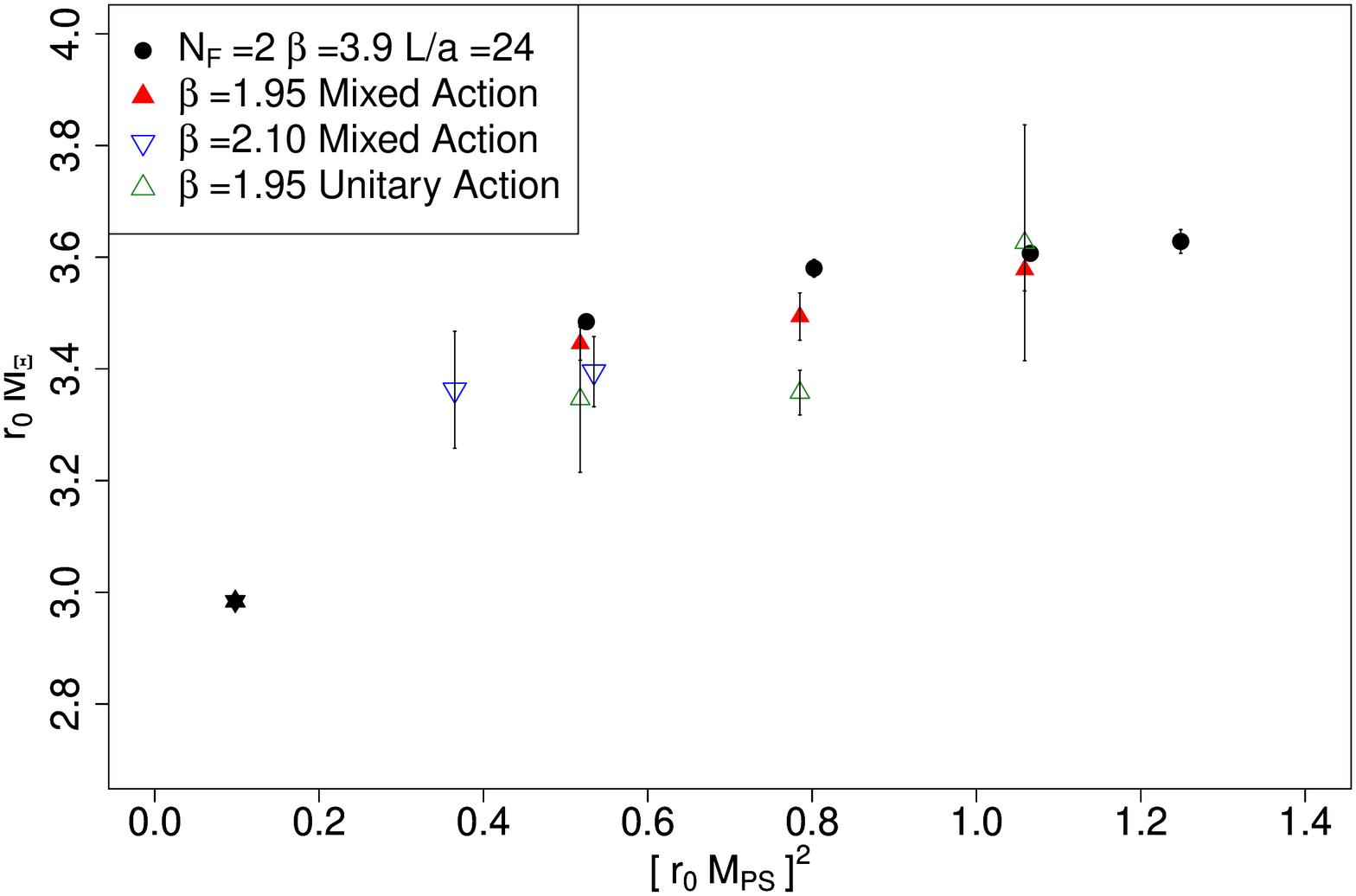} 
\vspace{-.4cm}
\caption{$\Xi$ mass   as a function of $(r_0 m_{\rm{PS}})^2$ both in the unitary and mixed action approach.For comparison $N_f=2$ data are shown with black dot.
The physical point is indicated by  a black star. We  used the value of $r_0$ from Ref.~\cite{Baron:2010bv} to set the scale.}
\label{fig:r0Xi}
\end{minipage}
\end{figure}

In Fig.~\ref{fig:r0lambda} we show the $\Lambda$ mass, computed in the mixed action approach explained in \ref{se:setup}, in units of $r_0$ as a function of $(r_0 m_{\rm{PS}})^2$ for the two finest lattice spacings. We also show our $N_f=2$ results obtained  at $\beta=3.9$. The results agree within the errors. A quantitative comparison is  made difficult by the fact that the sea strange quark mass at $\beta=1.95$ and $\beta=2.10$ are not matched.  Note also that the valence strange quark mass used in our $N_f=2$ computation has not been estimated. With the existing data,  we are unable to obtain quantitative estimates of the discretisation effects in our $N_f=2+1+1$  results  or to highlight some ``quenching effect''  with respect to the $N_f=2$ case.

In Fig.~\ref{fig:r0Xi},  besides the mass  of the $\Xi$  as a function of $(r_0 m_{\rm{PS}})^2$ obtained in the mixed action approach and in the $N_f=2$ case, we show the results obtained in the unitary setup. We observe that there is a good agreement except for ensemble B55.32 where the unitary theory gives a lower value. The behaviour at this particular value is under study.

\section{Conclusion }

We have shown preleminary results concerning the ground state  baryon spectroscopy  using $N_f=2+1+1$ simulations. 
The nucleon mass shows small discretisation effects and a chiral behaviour similar to the one obtained in our $N_f=2$ study. 
To our present accuracy, the isospin breaking is compatible with zero both in the $\Delta$ and $\Xi$ case. 
We briefly described the two strategies followed to determine masses in the strange sector, namely the so called ``mixed'' and ``unitary''  cases. 
The results agree within errors, but it is stressed that with the present amount of informations quantitative statements on discretisation or quenching effects would be premature. In all case the results obtained with  the mixed approach are in agreement with those in  the unitary case, with one exception in the case of the $\Xi$ where further investigation is under way.

\section*{Acknowledgements}
This work was performed using  HPC resources from GENCI/IDRIS (Grant 2010-052271)
and CCIN2P3 in Lyon. 
 M.P.  acknowledges financial support by a Marie Curie European
Reintegration Grant of the 7th European Community Framework
Programme under contract number PERG05-GA-2009-249309.

\end{document}